\documentclass[aps,prl,twocolumn,showpacs,superscriptaddress]{revtex4-1}
\usepackage{amsmath,amssymb}
\usepackage{graphicx}

\begin{document}
\title{Comment on ``Comment on ``Integrability of the Rabi Model''"}

\author{Murray T. Batchelor}
\email[]{batchelor@cqu.edu.cn}
\affiliation{Centre for Modern Physics, Chongqing University, Chongqing 400044, China}
\affiliation{Department of Theoretical Physics,
Research School of Physics and Engineering, and Mathematical Sciences Institute, 
Australian National University, Canberra ACT 0200, Australia}
\author{Zi-Min Li}
\affiliation{Centre for Modern Physics, Chongqing University, Chongqing 400044, China}

\date{\today}

\maketitle

It has been claimed \cite{comment} that Braak's solution \cite{Braak1} for the energy spectrum of the Rabi model 
\begin{equation}
H=\omega a^\dag a +g\sigma_z(a+a^\dag) +\Delta \sigma_x
\label{rabi}
\end{equation}
is not complete. The claim is incorrect.
Braak's solution \cite{Braak1,Braak2} is given by $E_n = x_n - g^2/\omega$, where $x_n$  is the $n$th zero of 
the functions 
\begin{equation}
G_\pm(x) = \sum_{n=0}^\infty K_n (x)  \left[ 1 \mp \frac{\Delta}{x - n\omega} \right] \left( \frac{g}{\omega} \right)^n .
\end{equation}
The coefficients $K_n(x)$ is defined recursively by
$n K_n = f_{n-1}(x) \, K_{n-1}  - K_{n-2}$, with 
\begin{equation}
f_n(x)  = \frac{2g}{\omega} + \frac{1}{2g} \left( n \omega - x  + \frac{\Delta^2}{x - n\, \omega } \right) ,
\end{equation}
and initial conditions $K_0=1, K_1(x)=f_0(x)$.

The exceptional part of the energy spectrum is defined by the eigenvalues 
$E_n=n\omega-g^2/\omega$, i.e., those eigenvalues with $x_n = n \omega$. 
Braak has shown that the condition $K_n(n\omega)=0$ gives the doubly degenerate 
eigenvalues found by Judd \cite{Judd}.
This condition arises in a natural way from the pole structure in Braak's solution \cite{Braak1}.
In the Comment \cite{comment} it was argued that the set of Juddian solutions is just a subset of the exceptional eigenvalues.
Using a determinant method in the Bargmann space the authors obtained  
another infinite set of exceptional eigenvalues which are non-degenerate and 
pass through the points $g=0,\Delta=\pm(n+1),\pm(n+2),\ldots$ in the $\Delta-g$ plane.
It was claimed that these lines are neglected in Braak's solution and thus that the solution is not complete.

\begin{figure}[ht]
  \centering
\includegraphics[width=0.235\textwidth]{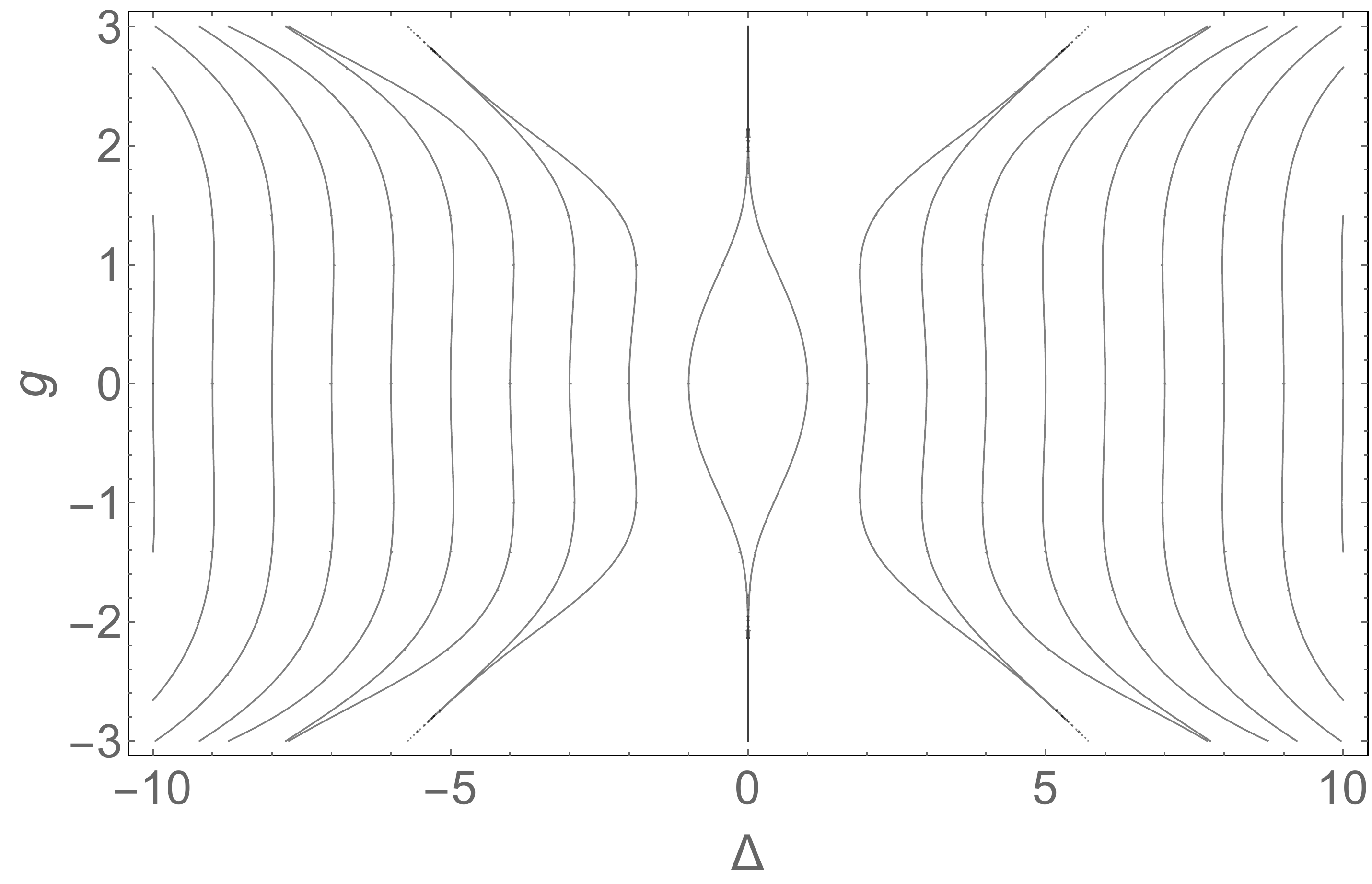}
\includegraphics[width=0.235\textwidth]{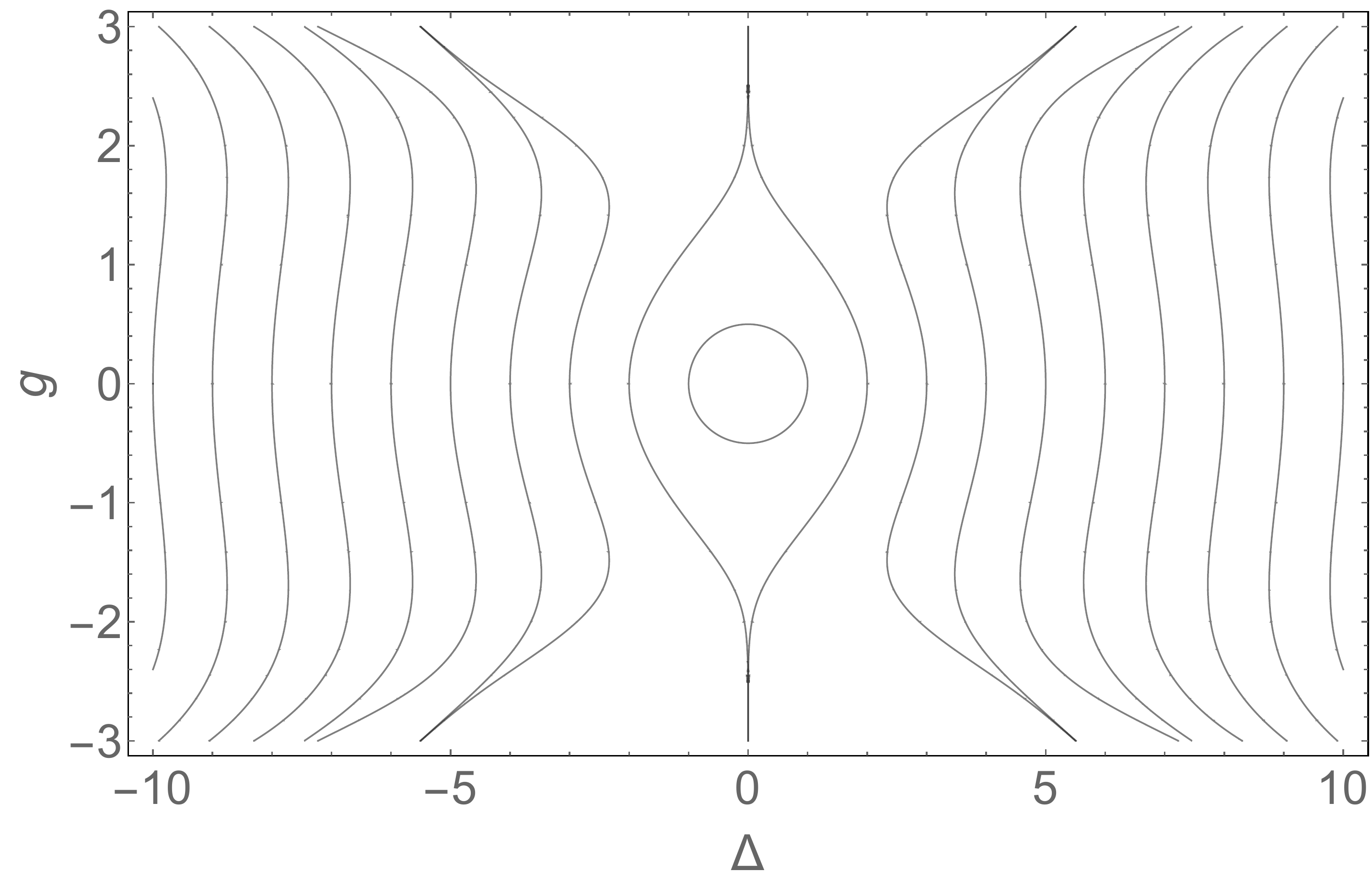}
\includegraphics[width=0.235\textwidth]{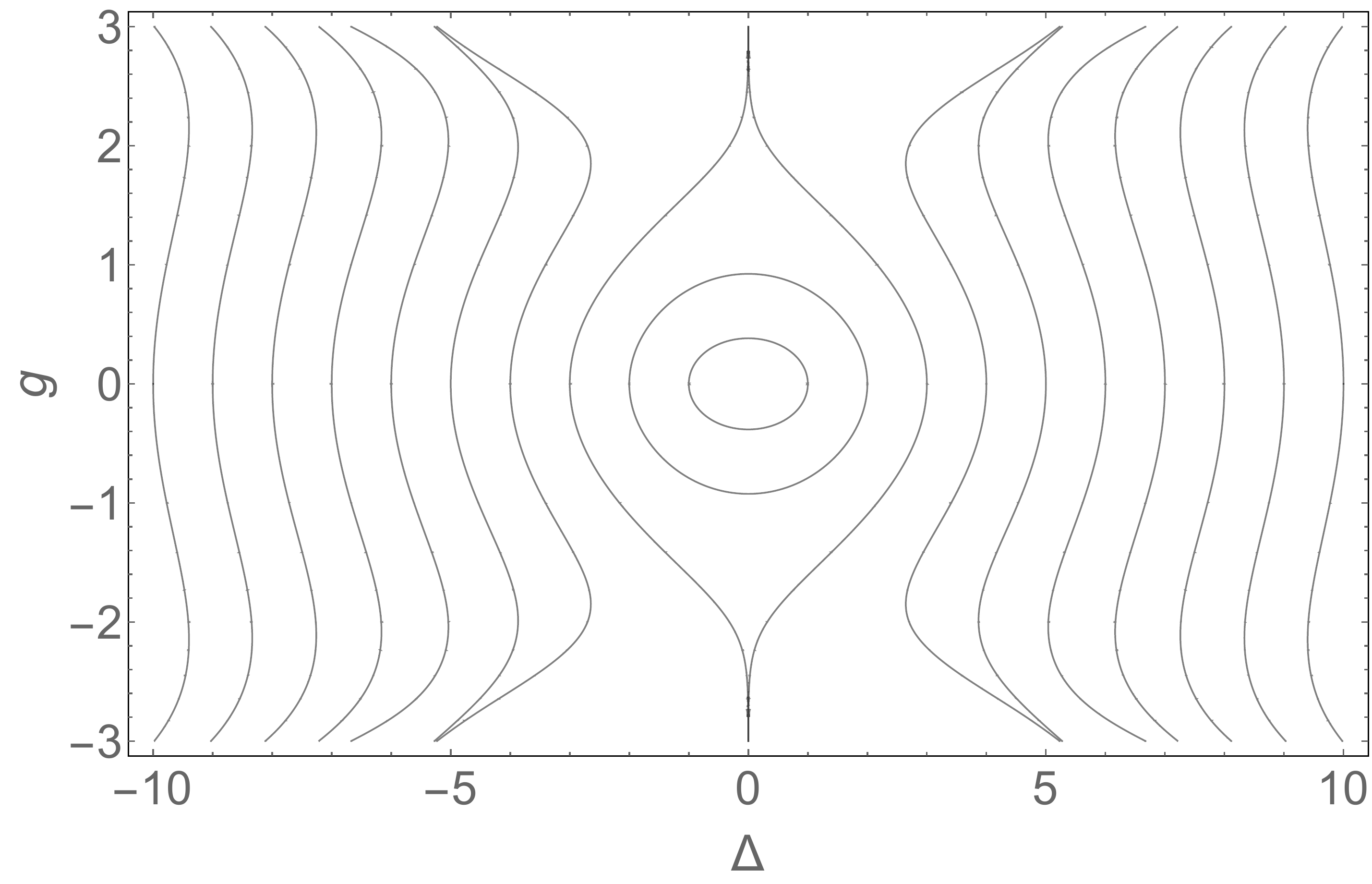}
\includegraphics[width=0.235\textwidth]{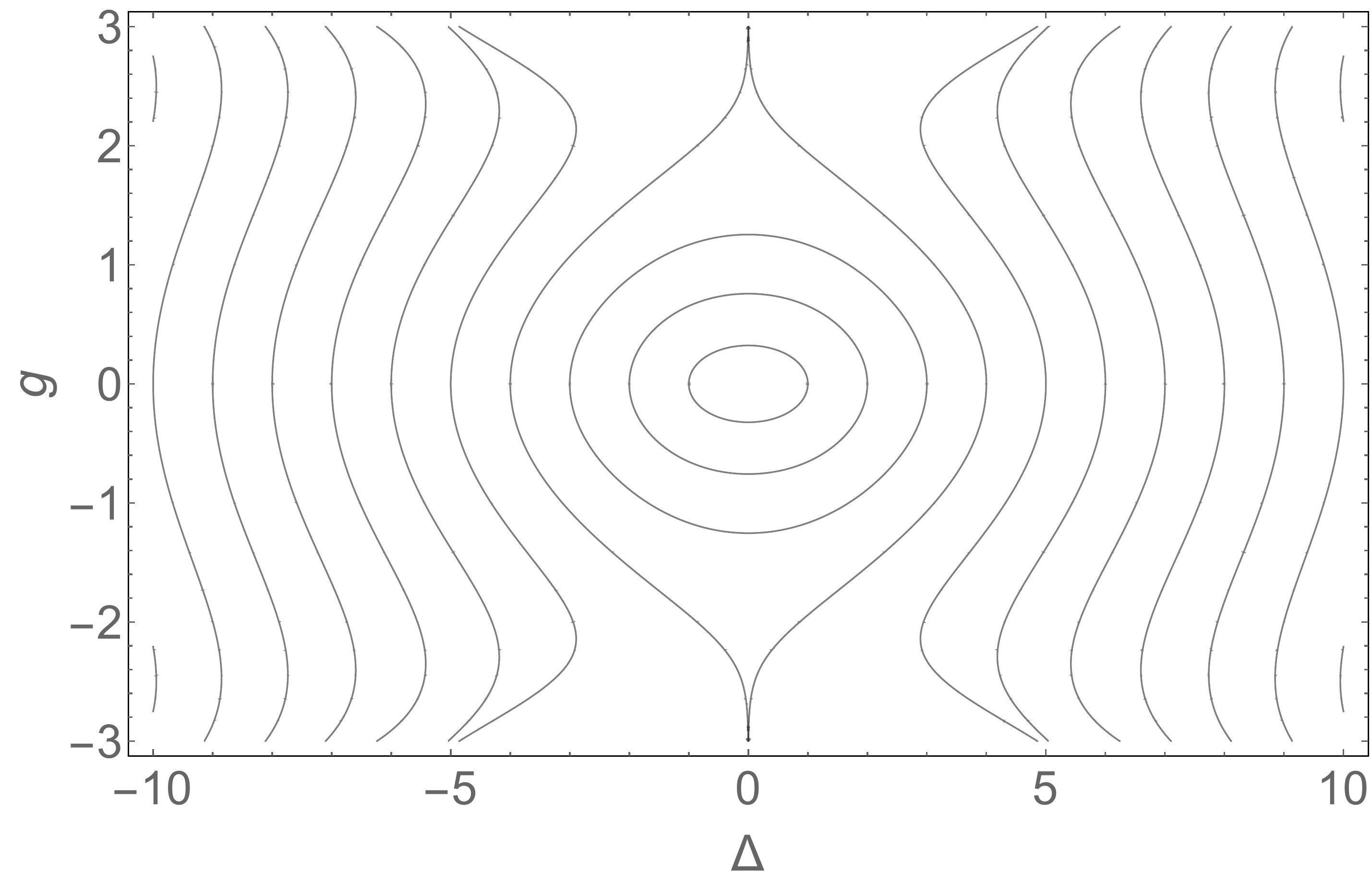}
\caption{Energy spectrum of the Rabi model for $n=0,1,2,3$ in the $\Delta$-$g$ plane obtained from Braak's solution. 
This figure is identical to that obtained in \cite{comment}.} 
\label{fig}
\end{figure}

However, this set of non-degenerate exceptional points has been discussed before (see, e.g., \cite{Braak2, MPS}).
In particular, Braak has shown  \cite{Braak2} that these points also arise naturally from lifting the pole structure, this time 
in one of the functions $G_-(x)$ or $G_+(x)$.
In Figure \ref{fig} we show plots of the energy spectrum in the $\Delta$-$g$ plane
obtained directly from the above solution.
These plots are identical to those given in \cite{comment}.
The $n$ closed loops correspond to the doubly degenerate Juddian points following from the condition  $K_n(n\omega)=0$. 
The other lines are the non-degenerate exceptional points also following directly from Braak's solution. 
The set of doubly degenerate Juddian points is particularly important as they have been 
shown to induce Dirac cones in the energy landscape of the driven Rabi model \cite{cones}.

In conclusion, it is already known that the set of doubly degenerate Juddian solutions is only a subset of all 
exceptional eigenvalues with the form $E_n=n\omega-g^2/\omega$ in the energy spectrum of the Rabi model.
The set of non-degenerate solutions of this form discussed in \cite{comment} also follows from Braak's solution.
Therefore the claim that Braak's solution is not complete is incorrect.

\end{document}